\documentclass[twocolumn,,amsmath,amssymb]{revtex4}
\usepackage{graphicx}% Include figure files
\usepackage{dcolumn}% Align table columns on decimal point
\usepackage{bm}% bold math
\newcommand{\etal}{\emph{et al.}}
\newcommand{\be}{\begin{equation}}
\newcommand{\ee}{\end{equation}}
\newcommand{\bfig}{\begin{figure}}
\newcommand{\efig}{\end{figure}}
\newcommand{\incl}{\includegraphics}
\begin{document}

\title{The thermopower and Nernst Effect in graphene in a magnetic field}

\author{Joseph G. Checkelsky and N. P. Ong
} \affiliation{ \mbox{Department of Physics, Princeton University,
New Jersey 08544, U.S.A.} }

\date{\today}
\begin{abstract}
We report measurements of the thermopower $S$ and Nernst signal
$S_{yx}$ in graphene in a magnetic field $H$. Both quantities show
strong quantum oscillations vs. the gate voltage $V_g$. Our
measurements for Landau Levels of index $n\ne 0$
are in quantitative agreement with the edge-current model of Girvin and
Jonson (GJ). The inferred off-diagonal thermoelectric conductivity
$\alpha_{yx}$ comes close to the quantum of Amps per Kelvin. At the
Dirac point ($n=0$), however, the width of the peak in $\alpha_{yx}$
is very narrow.  We discuss features of the thermoelectric response
at the Dirac point including the enhanced Nernst signal.
\end{abstract}

\maketitle                   % Produces the title

In graphene, the linear dispersion of the electronic states near the chemical potential $\mu$ is well described by the Dirac Hamiltonian.  As shown by Novoselov \etal~\cite{Novoselov1,Novoselov2,Novoselov3} and by Zhang \etal~\cite{Zhang1,Zhang2,Zhang3} quantization of the electronic states into Landau Levels leads to the integer quantum Hall Effect (QHE).  Because of the linear dispersion, the energy $E_n$ of the Landau Level (LL) of index $n$ varies as
$E_n = \mathrm{sgn}(n)\sqrt{2e\hbar v_F^2B|n|}$, where $B$ is the magnetic
induction, $v_F$ the Fermi velocity, $e$ the electron charge,
and $h$ is Planck's constant.  The quantized Hall conductivity is given by
\be
\sigma_{xy} = \frac{4e^2}{h}\left(n+\frac12\right),
\label{sxy}
\ee
where the factor 4 reflects the degeneracy $g$ of each LL
(2 each from spin and valley degrees).

%%%%%%%%%%%%%%%%%%%%%%%%%%%%%%%%%%%%%%%%
%%%%%%%%%%%%%%%%%%%%%%%%%%%%%%%%%%%%%%%%
%%%%%%%%%%%%%%%%%%%%%%%%%%%%%%%%%%%%%%%%
%%%%%%%%%%%%%%%%%%%%%%%%%%%%%%%%%%%%%%%% FIGURE
\bfig[h]            % Fig 1
\incl[width=7.5cm]{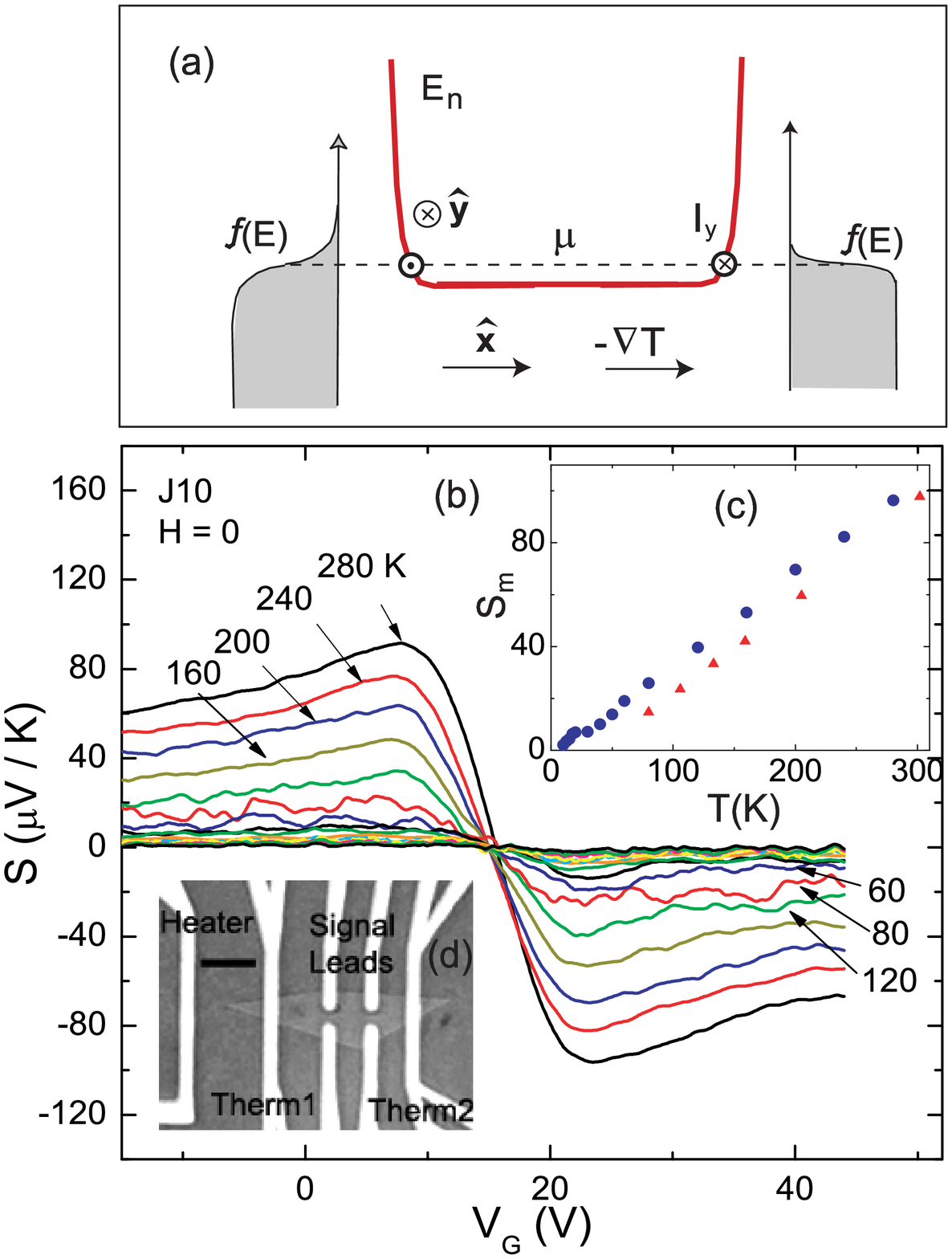} \caption{\label{figSvsV}
(Color online) (Panel a) The effect of $-\nabla T$ on the edge
currents $I_y$ in a QHE system ($n\ne 0$). The energy $E_n$ of a LL
(bold curve) increases very steeply at the sample edges, with $\mu$
the chemical potential (dashed line).  If $H_z>0$, $I_y$ is negative
(positive) at the left (right) edge, as indicated by open circles.
The magnitude $|I_y|$ is larger at the warmer edge. Fermi-Dirac
distributions $f(E)$ are sketched at the sides. (Panel b) Curves of
thermopower $S=-S_{xx}$ vs. gate voltage $V_g$ in Sample J10 at
selected $T$.  The curves are antisymmetric about the Dirac Point
which occurs at the offset voltage $V_0$ = 15.5 V. The peak value
$S_{m}$ is nominally linear in $T$ from 25 to 300 K (Panel c). Less
complete data from sample K59 are also plotted.  A photo of Sample
J10 (faint polygon) is shown in Panel (d).  A micro-heater as well
as thermometers (therm) and signal leads are patterned with
electron-beam lithography.  The black scale bar is 3 $\mu$m. } \efig

Detailed investigations of the longitudinal resistance $R_{xx}$ and
Hall resistance $R_{xy}$ have been reported by several groups
~\cite{Novoselov1,Novoselov2,Novoselov3,Zhang1,Zhang2,Zhang3,Check}.
By contrast, the thermoelectric tensor $S_{ij}$ is less
investigated.  $S_{ij}$ relates the observed electric field
$\bf E$ to an applied temperature gradient $-\nabla T$, viz.
${\bf E} = \stackrel{\leftrightarrow}{\mathbf S}\cdot(-\nabla T)$.
On the other hand,
the charge current density
$\bf J$ produced by $-\nabla T$ is expressed by
the thermolelectric conductivity tensor
$\stackrel{\leftrightarrow}{\alpha}$, viz.
${\bf J} =\; \stackrel{\leftrightarrow}{\alpha}\cdot(-\nabla T)$.
Although $\bf J$ is not measured directly, $\alpha_{ij}$ may be obtained by
measurements of both $S_{ij}$ and the resistivity
tensor $\rho_{ij}=R_{ij}$.
(By convention, $-S_{xx}$ is the thermopower $S$; we refer to
$S_{yx}= E_y/|\nabla T|$ as the Nernst signal.)

A most unusual feature of the thermoelectric response of
a QHE system (for $n\ne 0$) is that, despite the dominance of the off-diagonal
(Hall-like) current response, the thermopower displays a large peak
at each LL whereas the Nernst signal is small.
In the geometry treated by Girvin and Jonson~\cite{Girvin,Jonson}
(Fig. \ref{figSvsV}a), the 2D sample is of
finite width along $\bf\hat{x}$, but is
infinite along $\bf\hat{y}$ (with the applied magnetic field $\bf H||\hat{z}$).
As we approach either edge, the LL energy $E_n$
rises very steeply (bold curve).  At $T=0$, edge currents $I_y$ exist
at the intersections (open circles) of $E_n$ with the
chemical potential $\mu$. In a gradient $-\nabla T$, the
magnitude of $I_y$ is larger at the warmer edge than at the cooler edge
because of increased occupation of states above $\mu$.
The difference $|\delta I_y|$ is a maximum when $\mu$
is aligned with $E_n$ in the bulk.
The corresponding value of $\alpha_{yx}$ is then a universal
quantum $(k_Be/h)\ln 2$ with units of Amperes per Kelvin
($k_B$ is Boltzmann's constant).
In turn, $\delta I_y$ produces a quantized Hall
voltage $V_H = (h/e^2)\delta I_y$ that drops $||\bf\hat{x}$.
Hence, conflating these
2 large off-diagonal effects, the thermopower $S$ becomes
very large when $\mu$ aligns with $E_n$ ($n\ne 0$).
By contrast, the transverse (Nernst) voltage
is small (in the absence of disorder).

In graphene, this picture needs revision when $\mu$ is at
the Dirac point.  For the $n=0$ LL, the nature of the edge currents
is the subject of considerable
debate~\cite{Fertig,Abanin,Nomura,AbaninGeim}.
What are the profiles of $S$ and $S_{yx}$?  We have measured $S_{ij}$ and
$R_{ij}$ to a maximum $H$ of 14 T at 20 and 50 K.
Our results reveal that, at 9 T, the thermoelectric response in graphene
already falls in the quantum regime at 50 K.  The inferred
off-diagonal current response $\alpha_{xy}$ is a series of peaks
close to the quantum value $(gk_Be/h)\ln 2$ (independent of $n$, $B$ and $T$).
We compare our results with the caculations of Girvin and Jonson (GJ),
and discuss features specific to the $n=0$ LL at the Dirac point.

Kim and collaborators~\cite{Kim} have pioneered a
lithographic design for measuring the thermopower of
carbon nanotubes.  We have adopted their approach with
minor modifications for graphene.
Using electron-beam lithography, we deposited narrow gold lines
which serve as a micro-heater to produce $-\nabla T$ and
thermometers (Fig \ref{figSvsV}d).  The latter are
also used as current leads when $R_{xx}$ and $R_{xy}$ are measured.
Above 10 K, the thermometers can resolve $\delta T\sim \pm$1 mK.
The typical $\delta T$ is $\sim$10 mK between the 2 thermometers.
However, the small spacing of the voltage leads ($\sim$2 $\mu$m)
leads to an uncertainty of $\pm 10\%$ in estimating $\delta T$
between them.  A slowly oscillating current at frequency
$\omega$ is applied to the heater and the resulting thermoelectric
signals are detected at 2$\omega$ and $-90^{\circ}$ out of phase.

In our geometry with $-\nabla T||\bf\hat{x}$, $\bf H||\bf\hat{z}$,
the charge current density $\bf J$ is
(summation over repeated indices implied)
\be J_i =
\sigma_{ij}E_j + \alpha_{ij}(-\partial_j T)\quad (i = x, y).
\label{Jij}
\ee
The (2D) conductivity and thermoelectric conductivity
tensors are often written as $\sigma_{ij} = L^{11}_{ij}(e^2/T)$ and
$\alpha_{ij} = L^{12}_{ij}/T^2$, respectively. Setting $\bf J$ = 0,
we have for the observed $E$-fields
\be E_i =
-\rho_{ik}\alpha_{kj}(-\partial_j T) =
S_{ij}(-\partial_j T),
\label{Ei}
\ee
with $\rho_{ij}=R_{ij}$ the 2D resistivity tensor.
The thermopower $S = -E_x/|\nabla T|$
equals $\rho_{xx}\alpha_{xx}+\rho_{yx}\alpha_{xy}$
($S>0$ for hole doping), while the Nernst signal is given by
(with $\rho_{xx}=\rho_{yy}$)
\be S_{yx} =
\rho_{xx}\alpha_{xy}-\rho_{yx}\alpha_{xx}.
\label{Syx}
\ee
The 2 terms tend to cancel mutually, except at the Dirac point
where $\rho_{yx}$ vanishes (see below).

Inverting Eq. \ref{Ei}, we may calculate the tensor $\alpha_{ij}$
from measured quantities.  We have
\begin{eqnarray}
\alpha_{xx} &=& -(\sigma_{xx} E_x + \sigma_{xy} E_y)/|\nabla T| \nonumber\\
\alpha_{xy} &=& (-\sigma_{xy}E_x + \sigma_{xx} E_y)/|\nabla T|.
\label{aS}
\end{eqnarray}

Under field reversal ($\bf H\rightarrow$ -$\bf H$), $S$ is symmetric
whereas $S_{yx}$ is antisymmetric.  For each curve taken in field,
we repeat the measurement with $\bf H$ reversed.
All curves of $S$ and $S_{yx}$ reported here
have been (anti)symmetrized with respect to $\bf H$.
As for charge-inversion symmetry, we expect the sign of $S$ to change
with the shifted gate voltage $V_g'\equiv V_g-V_0$
(i.e. between hole and electron filling), but
the sign of $S_{yx}$ stays unchanged ($V_0$ is the offset voltage).
However, we have not imposed
charge-inversion symmetrization constraints on the curves.
Apart from field (anti)symmetrization, all the curves are the raw data.

%%%%%%%%%%%%%%%%%%%%%%%%%%%%%%%%%%%%%%%%
%%%%%%%%%%%%%%%%%%%%%%%%%%%%%%%%%%%%%%%%
%%%%%%%%%%%%%%%%%%%%%%%%%%%%%%%%%%%%%%%%
%%%%%%%%%%%%%%%%%%%%%%%%%%%%%%%%%%%%%%%% FIGURE

\bfig[t]            % Fig 2
\incl[width=9cm]{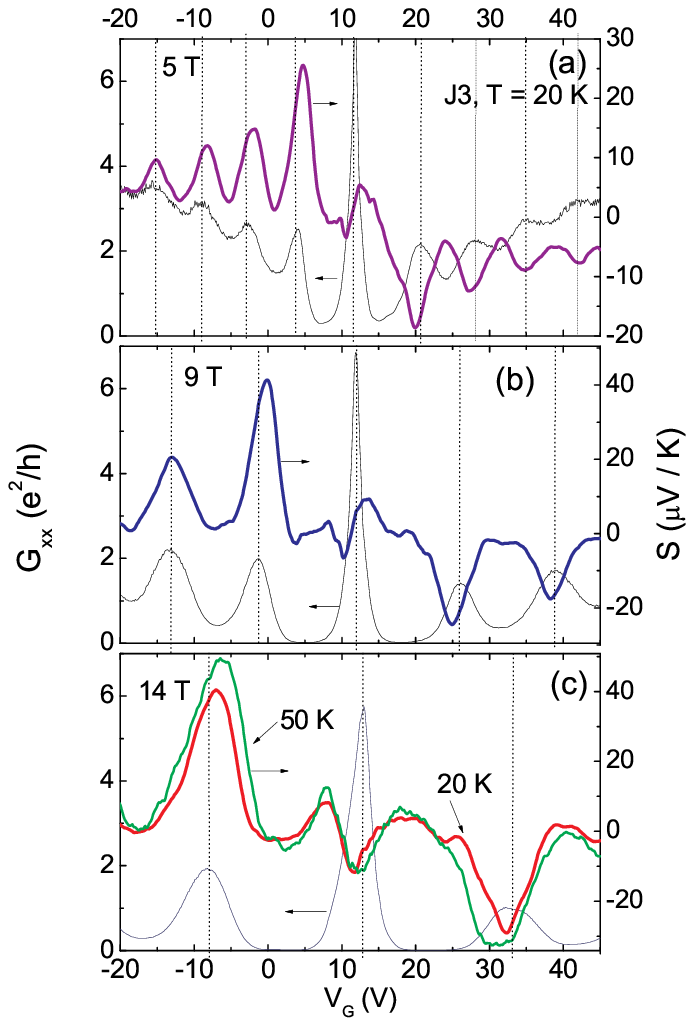} \caption{\label{figSG} (Color
online) Variation of thermopower $S$ vs. $V_g$ (bold curves) and
conductance $G_{xx}$ vs. $V_g$ (thin curves) in Sample J3 at $H$ =
5, 9 and 14 T (Panels a, b and c, respectively). The offset $V_0$ =
12.5 V.  All curves were measured at 20 K except for the curve at 50
K in Panel (c). Vertical lines locate the maxima of $G_{xx}$. }
\efig

Figure \ref{figSvsV}b shows traces of $S$ vs.
$V_g$ at selected $T$.  The thermopower $S$ changes sign as
$V_g$ crosses the charge-neutral point (Dirac Point),
assuming positive (negative) values on the hole (electron) side.
The peak value $S_{m}$ is nominally $T$-linear from $\sim$20 K
to 300 K (Fig. \ref{figSvsV}c).

%%%%%%%%%%%%%%%%%%%%%%%%%%%%%%%%%%%%%%%%
%%%%%%%%%%%%%%%%%%%%%%%%%%%%%%%%%%%%%%%%
%%%%%%%%%%%%%%%%%%%%%%%%%%%%%%%%%%%%%%%%
%%%%%%%%%%%%%%%%%%%%%%%%%%%%%%%%%%%%%%%% FIGURE

\bfig[t]            % Fig 3
\incl[width=9cm]{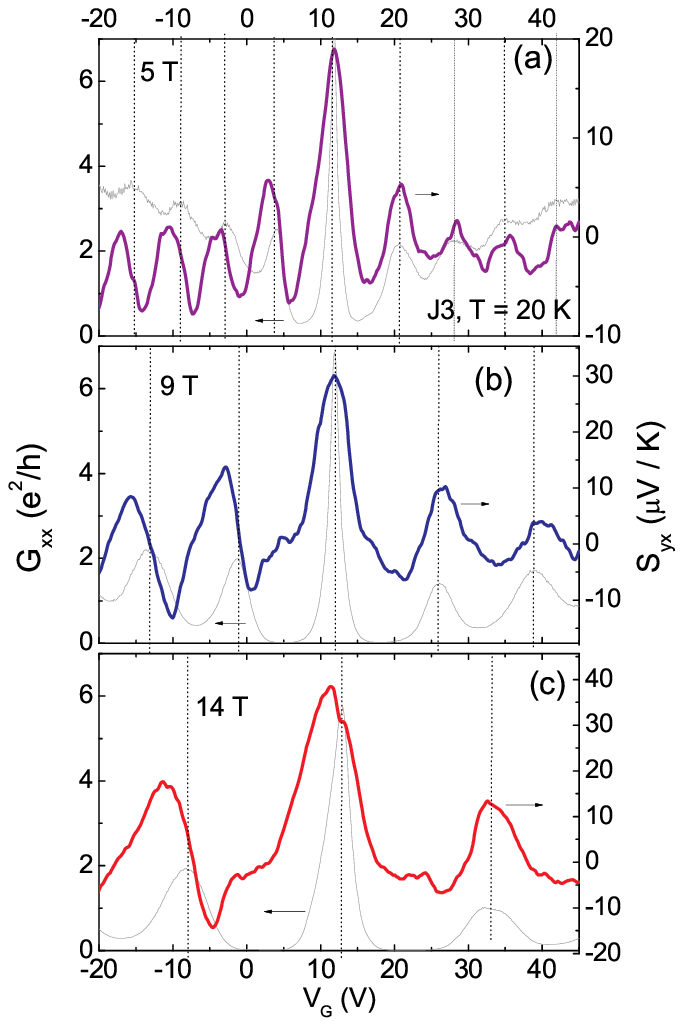} \caption{\label{figNernst}
(Color online) Variation of Nernst signal $S_{yx}$ vs. $V_g$ (bold
curves) and conductance $G_{xx}$ vs. $V_g$ (thin curves) in J3 at
the 3 field values $H$ = 5, 9 and 14 T (Panels a, b and c,
respectively). All curves were measured at 20 K. Vertical lines
locate the maxima of $G_{xx}$. The sign of $S_{yx}$ was incorrectly
assigned in a previous version of this paper~\cite{previous}. }
\efig

In sharp contrast to the smooth variation in Fig. \ref{figSvsV},
the curves of $S$ vs. $V_g$ show pronounced oscillations
when $H$ is finite, reflecting Landau quantization of the Dirac
states.  Figures \ref{figSG}a, b and c display $S$ vs. $V_g$ (bold curves) with $H$ fixed at the values
5, 9 and 14 T, respectively. For comparison, we
have also plotted (as thin curves) the corresponding conductance
$G_{xx}=\sigma_{xx}$.  (With the exception of the curve at 50 K in Panel (c),
all curves were measured at 20 K.)  Whereas at large $|n|$, the
peaks in $S$ are aligned with those in $G_{xx}$ (vertical lines),
at $n=\pm 1$, they disagree.  In Panel a, the peaks for $V_g'<0$ (hole doping)
decrease systematically in magnitude as $n$ increases from 1
to 4.

The curves of the Nernst signal $S_{yx}$
are displayed in Fig. \ref{figNernst}.  For $n\ne 0$, $S_{yx}$ displays
a dispersive profile centered at the vertical
lines, in contrast with the peak profiles of $S$. Moreover,
$S_{yx}$ is smaller in magnitude by a factor of 4-5.
For the $n$ = 0 LL, however, the profiles change character, with
$S_{yx}$ displaying a large positive peak.  The reason for
the enhancement of $S_{yx}$ at the Dirac point is
discussed below.  The positive sign of $S_{yx}$ at
the $n$=0 LL implies that the Nernst $E$-field
${\bf E}_N$ is parallel to ${\bf H}\times(-\nabla T)$~\cite{previous}.

%
%
%%%%%%%%%%%%%%%%%%%%%%%%%%%%%%%%%%%%%%%%
%%%%%%%%%%%%%%%%%%%%%%%%%%%%%%%%%%%%%%%%
%%%%%%%%%%%%%%%%%%%%%%%%%%%%%%%%%%%%%%%%
%%%%%%%%%%%%%%%%%%%%%%%%%%%%%%%%%%%%%%%% FIGURE

\bfig[t]            % Fig 4
\incl[width=9cm]{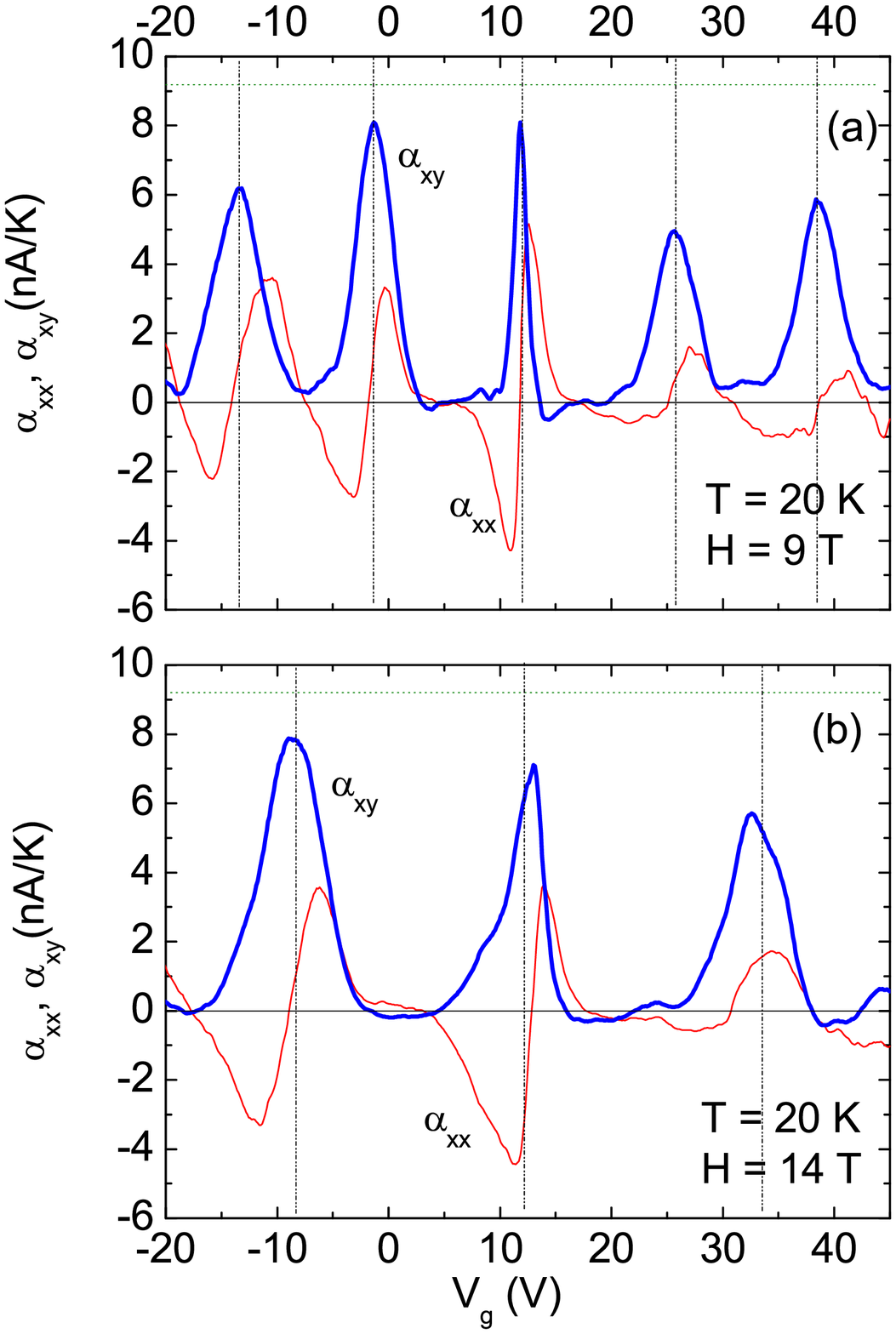} \caption{\label{figaxx} (Color
online) The thermoelectric response functions $\alpha_{xx}$ (faint
curve) and $\alpha_{xy}$ (bold) vs. $V_g$ at 9 T (Panel a) and 14 T
(b) at $T$ = 20 K calculated from $S$ and $S_{yx}$ (Eq. \ref{aS}).
In (a), the width of the peak at $n$=0 is narrower than the others
by a factor of 5. In both panels, vertical lines locate the peaks of
$G_{xx}$. The horizontal dashed line is $(4k_Be/h)\ln 2$. } \efig

In the theory of GJ~\cite{Girvin}, valid for GaAs-based devices, the
edge current difference is $\delta I_y = (ge/h)\sum_n
\int dk(\partial \epsilon/\partial k)(\partial f/\partial T)\delta T$,
with $f(\epsilon)$ the Fermi-Dirac distribution.
The off-diagonal term $\alpha_{xy}$ is given by
$(e/hT)\sum_n \int_{E_n}^{\infty}d\epsilon
(\epsilon-\mu) \left(-\frac{\partial f}{\partial\epsilon}\right).$
When $\mu=E_n$, $\alpha_{xy}$ attains a peak value,
corresponding to a quantized \emph{current} per Kelvin, given by ($g$=1)
\be
\alpha^{max}_{xy} =  \frac{k_Be}{h}\ln 2 \quad (\sim 2.32\;\mathrm{nA/K}).
\label{aQ}
\ee
GJ find that $S$ displays a series of peaks,
with the peak value at LL $n$ given by
\be S_{peak}(n) = \frac{k_B}{e}\frac{\ln 2}{(n+\frac12)}.
\label{Speak}
\ee
At low $T$, $S$ is independent of $H$
and $T$.

In Fig. \ref{figSG}, the peak value of $S$ at the $n=-1$ LL
increases from 25 at 5 T to 41 $\mu$V/K at both 9 and 14 T.
Moreover, as $T$ increases from 20 to 50 K (Panel c), the peak
increases only weakly (41 to 48 $\mu$V) in sharp contrast with the
$T$-linear behavior at $H$=0 (Fig. \ref{figSvsV}, inset). We thus
confirm the prediction of GJ that $S$ (at the peak) saturates to a
value independent of $T$ and $H$ at sufficiently low $T$. This
saturation contrasts with the $T$-linear behavior of $S$ in $H$=0
(Fig. \ref{figSvsV}c).

In graphene, however, Berry phase effects lead to a the
$\frac12$-integer shift in Eq. \ref{sxy}~\cite{Zhang1}.
In evaluating $\sigma_{xy}\sim\sum_n f(E_n)$, the $\frac12$-integer shift implies that
$S_{peak}(n)$ decreases as $k_B\ln 2/(en)$, instead of Eq. \ref{Speak}.
The measured values $S_{peak}$ = 41 $\mu$V/K at $n$ = -1 at 9 T
already exceeds slightly the predicted value 39.7 $\mu$V/K in Eq. \ref{Speak}.
Future experiments on cleaner samples may yield values closer to the
predicted value 59.6 $\mu$V/K for Dirac systems.
Regardless, for LL with $n\ne 0$, our observations are generally
consistent with the GJ theory. In principle, Eq. \ref{Speak} provides a way
to measure $\delta T$ on micron-scales with a resolution approaching voltage
measurements.

The most interesting question is the thermoelectric
response of the $n$=0 LL.  This is easier to analyze using
the pure thermoelectric currents $\alpha_{xx}$ and $\alpha_{xy}$ (obtained using
Eqs. \ref{aS}).  In Fig. \ref{figaxx}a, $\alpha_{xy}$ and ($\alpha_{xx}$)
is plotted as bold (thin) curves for $H$=9 T and $T$ = 20 K.
Panel (b) shows the curves at 14 T.  Compared with $S$ vs. $V_g$,
the peaks in $\alpha_{xy}$ are much narrower and clearly separated
by intervals in which $\alpha_{xy}$ is nominally zero.
Likewise, the purely dispersive
profile of $\alpha_{xx}$ is also more apparent.  Consistent with
the GJ theory, the overall magnitude of $\alpha_{xy}$ (for $n\ne 0$)
is larger than that of $\alpha_{xx}$.

A striking feature of $\alpha_{xy}$ is that its peaks are independent
of $n$.  Their average value $\sim$75 nA/K reaches to within 30$\%$ of the quantized
value of Eq. \ref{aQ} with $g=4$ (horizontal dashed line).  The largest uncertainty
in our measurement is in estimating the gradient between the voltage leads.
For the $n$ = 0 LL, the shortfall may also reflect incipient splitting
of the Landau sublevels (compare 9 and 14 T traces).
The uniformity of
the peaks in $\alpha_{xy}$ accounts for the observed enhancement of the Nernst
peak at the Dirac point.  By Eq. \ref{Syx}, $S_{yx}$ is the difference of 2
positive terms. For $n\ne 0$, the 2 terms are matched, and the partial
cancellation leads to a dispersive profile.  However, for $n=0$,
the vanishing of $\rho_{yx}$ as $V_g'\rightarrow 0$ strongly suppresses
the second term $-\rho_{yx}\alpha_{xx}$.  The remaining
term $\rho_{xx}\alpha_{xy}$ then dictates the size and profile of the Nernst peak.

We note that the sign of $\alpha_{xy}$ is a direct consequence of
the edge-currents~\cite{Girvin}.  When $H_z>0$,
$I_y<$0 on the warmer edge and $I_y<$0 on the cooler edge,
as depicted in Fig. \ref{figSvsV}a.  As a result, $\alpha_{xy}>0$
for both hole-and electron-doping.

Our measurements are consistent with the GJ theory for $n\ne$0.
For the $n$=0 LL, however, there is considerable uncertainty about
the nature of the edge states in graphene (or whether they exist in large $H$).
The simple edge-current picture for understanding the peaks in $\alpha_{xy}$
may need significant revision for $n=0$, despite the similarity of the peak
magnitude.  We also note that the peak at $n$ =0 is
much narrower (by a factor of 5) than the other peaks.  This is
also not understood.  By going to cleaner samples
and higher fields, we hope to exploit this narrow width
to resolve splitting of the 4 sublevels at $n$ = 0.

We are grateful to P. A. Lee for many discussions.
The research is supported by NSF through a MRSEC grant (DMR-0819860).

\emph{Note added}  After we completed these experiments, we learned
of 2 thermopower and Nernst experiments on graphene posted recently
(Yuri M. Zuev \etal, cond-mat arXiv: 0812.1393  and
Peng Wei \etal, cond-mat arXiv: 0812.1411).

%%%%%%%%%%%%%%%%%
%%%%%%%%%%%%%%%%%
%%%%%%%%%%%%%%%%%

%

%
\end{document}